\documentclass[journal=jacsat,manuscript=article]{achemso}

\usepackage[version=3]{mhchem} 

\author{Bruno Ipaves}
\affiliation{Instituto de F\'{\i}sica, Universidade de S\~ao Paulo, S\~ao Paulo, SP, Brazil}
\author{Jo\~ao F. Justo}
\affiliation{Escola Polit\'ecnica, Universidade de S\~ao Paulo, S\~ao Paulo, SP, Brazil}
\author{Lucy V. C. Assali}
\affiliation{Instituto de F\'{\i}sica, Universidade de S\~ao Paulo, S\~ao Paulo, SP, Brazil}
\email{ipaves@if.usp.br}

\title[An \textsf{achemso} demo]
  {Aluminum functionalized silicene: a potential anode material for alkali metal ion batteries}

\abbreviations{IR,NMR,UV}

\begin{document}







\begin{abstract}
 We have investigated the possibility of using aluminum functionalized silicene trilayers (\ce{ABC-Si4Al2}) as an anode material for alkali metal ion batteries (AMIBs). First, we studied the thermodynamic stability of \ce{ABC-Si4Al2} using {\it ab-initio} molecular dynamics simulations, showing that this material remains stable up to 600 K. Then, we explored the properties of alkali metal atoms (Li, Na, K) adsorption in \ce{ABC-Si4Al2}, finding several available sites with high adsorption energies. Moreover, we computed the diffusion properties of those atoms along high-symmetry paths using the nudged elastic band method. The results indicated diffusion barriers as low as those in graphite, especially for Na (0.32 eV) and K (0.22 eV), which allows those ions to migrate easily on the material's surface. Our studies also revealed that the full loaded \ce{Li4Si4Al2}, \ce{Na2Si4Al2}, and \ce{K2Si4Al2} systems provide low open-circuit voltage, ranging from 0.14 to 0.49 V, and large theoretical capacity of 645 mAh/g for Li- and 322 mAh/g for Na- and K-ion batteries, values that are close to the ones in other anode materials, such as graphite, \ce{TiO2}, and silicene-based systems. Those results indicate that aluminum functionalized few-layer silicene is a promising material for AMIBs anodes, particularly for Na- and K-ion batteries.
\end{abstract}

\section{Introduction}

The current challenges of energy transition toward renewable sources, such as solar and wind power systems, have been intrinsically associated with the search for efficient technologies for energy storage. Considering energy storage in the electric form, the lithium-ion battery (LIB) has been a leading option, which has been integrated into several electronic devices since entering the market around thirty years ago \cite{ma20212021}. 

Although the LIB technology has been widely used, it still carries some limitations when applied to large-scale systems, such as slow charging and degradation after only a few hundred cycles \cite{pomerantseva2017two}. Moreover, lithium is not an abundant chemical element on Earth \cite{hwang2017sodium}. Therefore, it is crucial to investigate alternative materials for batteries, in particular, new materials for electrodes. This represents an essential step toward improving the performance of rechargeable batteries to meet the world's increasing demand for renewable energies.

Sodium-ion (SIBs) and potassium-ion (KIBs) batteries have received attention in recent years since they possess working principles, and physical and chemical properties similar to LIBs. Moreover, the sodium (Na) and potassium (K) chemical elements are more abundant and less expensive than lithium (Li) \cite{rajagopalan2020advancements, tapia20212021}. On the other hand, the larger Na and K ions, as compared to Li, represents a major challenge for the SIBs and KIBs, once they present high diffusion barriers, high lattice strain, and irreversible structural degradation \cite{ma20212021}. For example, in LIBs the anode is usually made of graphite, having a theoretical capacity of 372 mAh/g by forming \ce{LiC6} \cite{schweidler2018volume}. However, this value is reduced to 279 mAh/g when the K ion is used (\ce{KC8}) \cite{jian2015carbon}. Additionally, due to a thermodynamic instability in interpolating Na ions, graphite is rarely used as an anode in SIBs \cite{hwang2017sodium, tapia20212021}.

Within this context, two-dimensional (2D) materials have emerged as potential components of alkali metal ion batteries (AMIBs). These materials carry suitable properties for those applications, such as large surface-to-volume ratio and elasticity, which allows efficient transport and superior ion adsorption, as well as small volume expansion of the electrode materials \cite{tan2017recent, chen2020transition}. Among several two-dimensional materials investigated, since the discovery of graphene, silicene has emerged as a promising candidate for applications in batteries. Silicene is made of silicon, one of the most abundant chemical elements on Earth, that is also environmentally friendly \cite{an2020recent}. Furthermore, it has been determined theoretically that the silicene monolayer has a 954 mAh/g capacity for LIBs and SIBs \cite{tritsaris2013adsorption, zhu2016silicene},  larger than that of graphite. Correspondingly, experimental investigations have shown that crystalline silicene (c-silicene) exhibits great performance as an anode in LIBs and KIBs with storage capacities respectively of 721 mAh/g \cite{liu2018few} and 180 mAh/g \cite{sun2022reversible}.   

Several strategies have been explored to get better materials for AMIB electrodes, such as the chemical doping and composition, in order to improve the ion conductivity and diffusibility within the solid \cite{shao2019recent}. For example, it has been shown that \ce{Si3C}, a silicene-based compound in which some Si atoms have been replaced by C ones, could be a potential anode material for AMIBs \cite{wang2020ab}. Recently, other investigations have shown that \ce{TiS2} doped with B, C, N, O, F, or P atoms could be used as anodes, improving the overall performance of AMIBs \cite{tian2021adsorption}. Additionally,  theoretical investigations of the \ce{BSi3} and \ce{AlSi3} systems, in which the silicene monolayer was doped with B and Al atoms, respectively, have shown that they have better properties for Na- and K-ion batteries, when compared to pristine silicene \cite{zhu2018potential}. Therefore, doping 2D systems with different chemical elements represents a promising strategy to increase the performance of the AMIBs. 

The physical properties of functionalized few-layer silicene compounds, specifically bilayers (\ce{Si2X2}) and trilayers  (\ce{Si4X2}) (with X = B, N, Al, P), have recently been explored by us using first-principles investigations \cite{ipaves2022functionalized}, showing that many functionalizations are energetically favorable, displaying large negative enthalpy of formation. The resulting compounds have shown metallic or semiconducting behavior, indicating several potential applications. In particular, the systems with metallic behavior, such as silicene functionalized with Al, has properties that make them good candidates to work as battery electrodes. The aluminum doping enlarges the lattice parameter of the original silicene material by about 7\% \cite{ipaves2022functionalized}, which may increase the adsorption of Li-, Na-, and K-ions, which is one of the requirements for superior anodes \cite{zhangchallenges}. 

Here, we performed a theoretical investigation on the properties of aluminum-doped silicene trilayer \ce{ABC-Si4Al2}, and its potential application as an AMIB anode. First, we explored the thermodynamic stability of this system as a function of temperature, showing that \ce{ABC-Si4Al2} remains structurally stable up to  600 K. Since the rate performance of AMIBs is associated with ionic diffusion properties \cite{shao2019recent}, we then calculated the adsorption energies of Li, Na, and K atoms in several sites of \ce{ABC-Si4Al2}, as well as their diffusion barriers along several high-symmetry pathways. Finally, we investigated the effects of alkali metal concentration and estimated the open-circuit voltage and the specific theoretical capacity of \ce{ABC-Si4Al2} as an anode material for AMIBs. 
The results indicated that \ce{ABC-Si4Al2} provides migration energy barriers and open-circuit voltage as low as in commercial anode materials, such as graphite. Moreover, the \ce{ABC-Si4Al2} maintains a good capacity for lithium, which is similar to that of pristine crystalline silicene and larger than that of graphite. Interestingly, the capacity for potassium is enhanced when compared to that of pristine crystalline silicene, and for sodium, we found the same capacity, showing that aluminum functionalized silicene trilayers (\ce{ABC-Si4Al2}) is a promising anode material for AMIBs.   

{\section{Computational  Details}}

The first-principles calculations were performed within the framework of the density functional theory \cite{hohenberg1964inhomogeneous,kohn1965self}, with the Quantum ESPRESSO computational package  \cite{Giannozzi2009,giannozzi2017advanced}. We used the exchange-correlation potential based on the generalized gradient approximation to describe the electronic interations, as proposed by Perdew, Burke, and Ernzerhof (PBE) \cite{perdew1996generalized}. In order to properly describe the effects of the dispersive van der Waals interaction \cite{zhang2011van, park2015van, marcondes2018importance}, the Dion {\it et al.} scheme \cite{dion2004}, optimized by Klime{\v{s}} {\it et al.} (optB88-vdW) \cite{klimevs2009}, was used. 

The projector augmented-wave (PAW) method \cite{Kresse} was applied to describe the interactions between ions and electrons,  with a cutoff energy of 1100~eV. To compute the electronic states, the Brillouin zones were sampled by a $16 \times 16 \times 1$ Monkhorst-Pack $k$-point grid \cite{Monkhorst}. Self-consistent iterations were performed until the total energy difference between two consecutive steps was lower than 0.1 meV/atom, while the structural optimization was performed until forces were lower than 1 meV/{\AA} in any ion, where relaxations and distortions were considered in all ions, without symmetry constraints.

The 2D structures were built with hexagonal simulation supercells, considering periodic boundary conditions. We used a lattice parameter of 25 {\AA} in the perpendicular direction to the nanosheet ($z$-axis), which is large enough to avoid interactions among cell images in that direction. The methodology presented so far has already been used in several investigations, providing an appropriate description of similar 2D systems \cite{ipaves2019carbon, ipaves2022functionalized}. 

 The adsorption energies ($E_{ads}$) and the diffusion barriers of the M = Li, Na, or K alkali adatoms were obtained by constructing a $3\times 3 \times 1$ supercell of the ABC-\ce{Si4Al2} system and, after the structural optimization, the adsorption energies were calculated by the equation 
\begin{equation}
E_{ads} ({\rm{M}}) = \Big[E_t\big({\ce{Si4Al2M}_n}\big) - E_t\big({\ce{Si4Al2}}\big) - nE_t\big({\rm M}\big)\Big]\Big/n\,, 
 \label{eq_E_ads}
\end{equation}
\noindent
where $E_t\big({\ce{Si4Al2M}_n}\big)$ is the total energy of the host nanosheet with $n$M alkali adatoms, $E_t\big({\ce{Si4Al2}}\big)$ is the total energy of an isolated ABC-\ce{Si4Al2} trilayer, and $E_t\big({\rm M}\big)$ is the total energy, per atom, of the M alkali metal in a body-centered cubic crystalline structure, computed within the same methodology described previously.  

For the diffusion barriers, we used the nudged elastic band (NEB) method \cite{henkelman2002methods} as implemented in the Quantum ESPRESSO package \cite{giannozzi2017advanced}. Two adjacent sites with the largest negative $E_{ads}$ were chosen as initial and final configuration reference states, where seven images were considered along the minimum energy pathway (MEP). 

The open-circuit voltage (OCV) is related to the $E_{ads}$ by
\begin{equation}
{\rm OCV} = -\frac{\Delta G_{\!f}}{ne} \approx - \frac{E_{ads}}{ne},
 \label{eq_ocv}
\end{equation}
\noindent
where $\displaystyle \Delta G_{\!f} = E_{ads} + P\Delta V_{\!f} - T\Delta S_{\!f}$ is the Gibbs free energy and, at room temperature, $\displaystyle P \Delta V_{\!f}$ and $ \displaystyle T\Delta S_{\!f}$ can be neglected \cite{he2019density}. Therefore, OCV was obtained using equation (\ref{eq_ocv}) by computing $E_{ads}$ as defined in equation (\ref{eq_E_ads}), when inserting adatoms into  \ce{ABC-Si4Al2} trilayer.

The theoretical specific capacity ($C$) was obtained by using 
\begin{equation}
{C} = \frac{n_{max} \times z \times {\rm F} \times 10^{3}}{A_{\ce{Si4Al2}}},
\label{eq_capacity}
\end{equation}
\noindent
where $\rm{F}=26.810$~Ah/mol is the Faraday constant, $z=1$ is the charge of the intercalating ion (Li, Na, or K), $n_{max}$ is the maximum number of M alkali adatoms that the host can store, and $\displaystyle A_{\ce{Si4Al2}}$ is the atomic mass of the host. 

Additionally, we used {\it ab-initio} molecular dynamics (AIMD) simulations, as implemented in the Vienna {\it ab-initio} simulation package (VASP) \cite{kresse1996efficient}, to explore the thermodynamic stability of the \ce{ABC-Si4Al2} host system. To allow possible structural reconstructions, a $5\times 5 \times 1$ supercell,  with 150 atoms, was used. The time step of the simulations was set to 1 fs and a Nose–Hoover thermostat (NVT) ensemble was employed to carry out the calculations at several temperatures (from 0 to 1000 K) for 5 ps.

\section{Results}

Silicene bilayers and trilayers, with several surface functionalizations (atomic species B, N, Al, P) and stacking configurations, were recently investigated by us using first-principles calculations \cite{ipaves2022functionalized}. According to the phonon spectra results and Born stability criteria, only some of the functionalized systems were found to be mechanically stable. In addition, most of these stable structures exhibit insulating or semiconducting electronic behavior, which is not entirely suitable for applications associated with battery electrodes.

\begin{figure}[htb]
\centering
\includegraphics[scale = 0.25, trim={0cm 0cm 0cm 0cm}, clip]{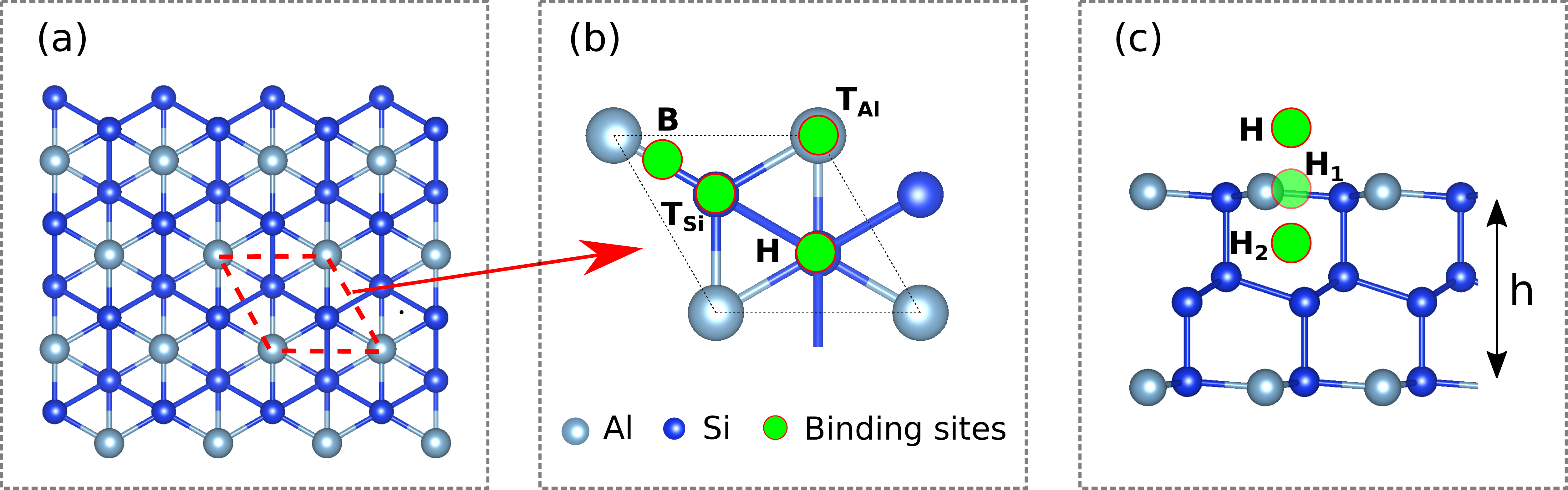} \quad
\caption{(a) Top view, (b) primitive cell, and (c) side view of the ABC-Si$_4$Al$_2$ and the binding sites for Li-, Na-, and K-alkali metal atoms. The hollow, top, and bridge sites are represented respectively by H, T, and B symbols. The dashed lines indicate the unit cell limits. The blue and grey spheres represent respectively Si and Al atoms, while the green spheres represent the binding sites for adsorption of the alkali metal ions.} 
\label{figure_1}
\end{figure}

On the other hand, among the stable systems studied, the silicene trilayer functionalized with aluminum atoms in the ABC stacking configuration, labeled as ABC-Si$_4$Al$_2$, is metallic and its lattice parameter is larger than the pristine few-layer silicene \cite{ipaves2022functionalized}, which could allow a large set of adsorption sites. Those properties already suggest that ABC-Si$_4$Al$_2$ system is a promising candidate for applications as battery electrodes. Figure \ref{figure_1} presents a schematic representation of the ABC-Si$_4$Al$_2$ structure. The optimized lattice parameters of the primitive cell are $a$ = $b$ = 4.143 {\AA}, with an $h$ = 6 {\AA} thickness.  
 
\subsection{Thermodynamic Stability}

\begin{figure}[hbt]
\centering
\includegraphics[scale = 0.19, trim={0cm 0cm 0cm 0cm}, clip]{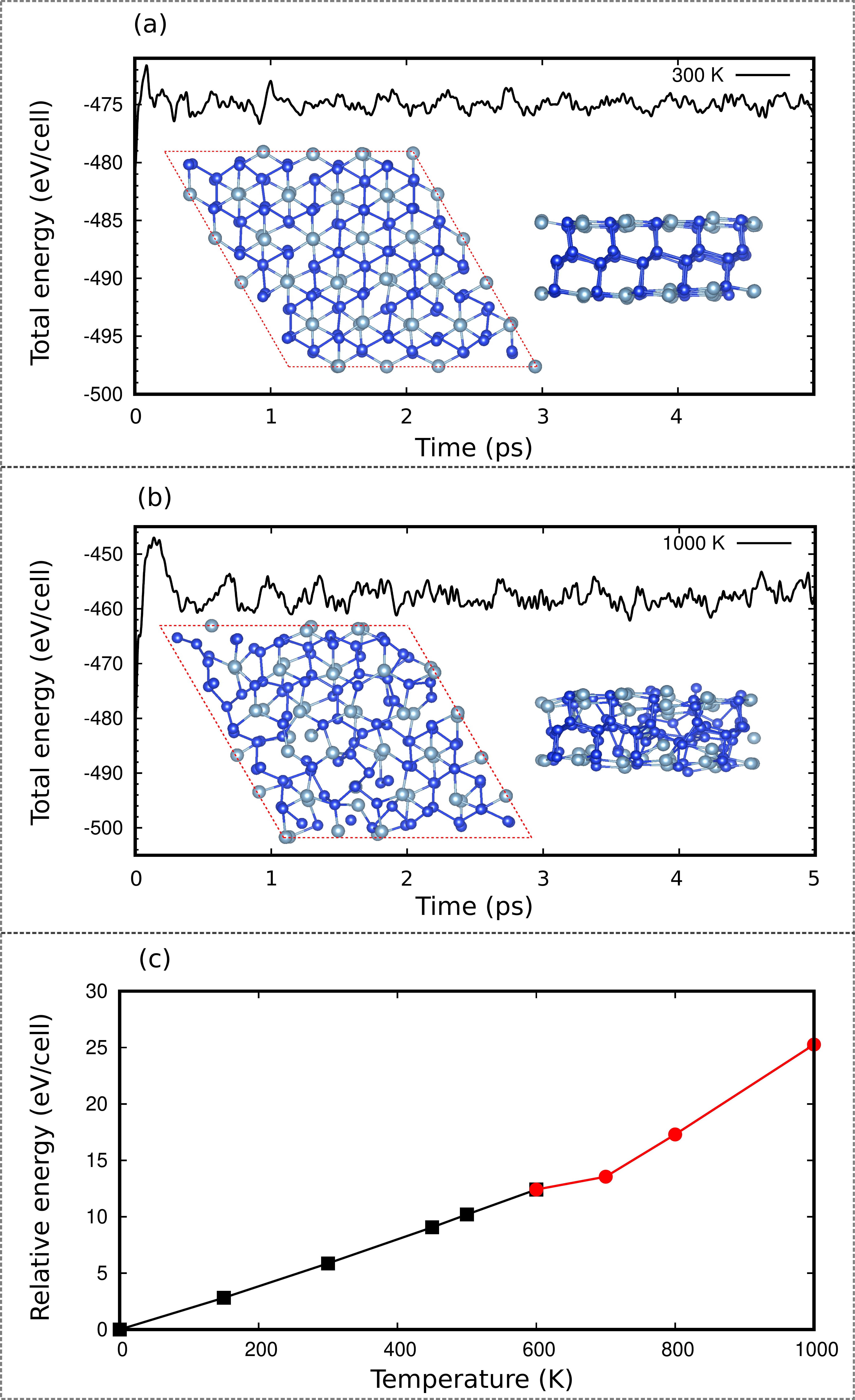} \quad
\caption{Total energy over 5 ps during AIMD simulations at (a) 300 K and (b) 1000 K, along with the respective snapshots of the final structures (top and side views) at the end of the simulation; (c) variation in the average total energy, relative to the reference value at T = 0 K, of the last 2 ps of the simulations as a function of temperature. The coloring of spheres in (a) and (b) is consistent with those in figure \ref{figure_1}.}
\label{figure_2}
\end{figure}

The thermodynamic stability of the ABC-Si$_4$Al$_2$ structure was investigated with AIMD simulations, using a 150-atoms supercell and lattice parameters of $a$ = $b$ = 20.715  {\AA}, at several temperatures from 0 to 1000 K. Figure \ref{figure_2}(a) shows the total energy of the system at 300 K during 5 ps. After about 1 ps, the system reaches thermal equilibrium, and then the energy oscillates around an equilibrium value. The figure also shows a snapshot of the final structure, indicating that the system keeps its structural integrity, i.e., the atoms only oscillate around their crystalline equilibrium positions, with no broken bonds or defect formation. Similar behavior is observed at temperatures up to 600 K. At higher temperatures, there is a clear change in the behavior of the system, as it starts to present broken bonds. Figure \ref{figure_2}(b) shows the total energy of the system at 1000 K during 5 ps, as well as a snapshot of the final structure, pointing out structural degradation, with a large concentration of broken bonds.

By inspecting the final structure in the simulation, it is already possible to estimate that 600 K is the temperature threshold at which the system can maintain its structural integrity. Below this threshold, the atoms essentially oscillate around their crystalline equilibrium positions, without breaking bonds, while over this threshold, some broken bonds could be observed. In order to determine this threshold more precisely, we investigated the behavior of the average total energy of the system in equilibrium as a function of the temperature. Figure \ref{figure_2}(c) shows the average total energy of the last 2 ps of the simulations as a function of temperature. It can be observed that the average total energy increases linearly with temperature below 600 K and,  taking into account the virial theorem, it is clear that the average kinetic and potential energies are very close, indicating structural integrity. Over 600 K, figure \ref{figure_2}(c) shows that the average total energy still increases linearly with temperature, but faster,  pointing out that there are broken bonds and defects, as well as some atomic diffusion. 

Those results demonstrate that the ABC-Si$_4$Al$_2$  structure is considerably stable for temperatures up to 600 K, and could be used as electrodes for applications below that threshold. Nevertheless, the nanosheet suffers considerable structural degradation at higher temperatures, and close to 1000 K, they are unsuitable for those applications.

\subsection{Ion Adsorption Properties}

In the previous section, we showed that the \ce{ABC-Si4Al2} compound keeps its structural stability up to 600 K. We now explore the possibility of using this material as an anode in AMIBs. For that, two important properties were investigated: how the alkali metal ions (Li, Na, or K) are adsorbed on the host surface, as well as incorporated inside the \ce{ABC-Si4Al2} system, and how they diffuse within the surface. We used a $3 \times 3 \times 1$ supercell, with optimized lattice parameters of $a$ = $b$ = 12.429~{\AA}, which corresponds to a chemical stoichiometry of \ce{ABC-M1Si36Al18} (M=Li, Na, K). This supercell is large enough for the simulation of one alkali atom adsorbed/inserted in the system, while avoiding interaction among images.

We first computed the adsorption energies of alkali metals in several high-symmetry binding sites on the trilayer surface, as shown in figure \ref{figure_1} (b), and subsequently in two sites inside the trilayer structure, as presented in figure \ref{figure_1} (c). 
We considered the hollow site (H) at the center of a hexagon, the top site (T) directly above each atom, and the bridge site (B) at the midpoint of a Si-Al bond. The bulk interstitial adsorption was also considered in two sites: the layer hollow site (H$_{1}$), at the center of a hexagonal ring, and the interstitial hollow site (H$_{2}$), inside the structure at the center of a hexagon, as presented in figure \ref{figure_1} (c).

After performing structural optimization, the alkali atoms moved from the B to the T$_{\rm Al}$ site and from the H$_{1}$ to the H$_{2}$ one, indicating that the ions are unlikely to be adsorbed at the B and H$_{1}$ sites. Therefore,  using equation (\ref{eq_E_ads}), we calculated the adsorption energy (E$_{ads}$) at the remaining sites, which are presented in  Table \ref{table_energy} for Li-, Na-, and K-adatoms in the H, T, and H$_{2}$ sites. The negative values of E$_{ads}$ indicate that the sites are energetically favorable for adsorption. The values in the table are in the same range as those in silicene-based materials, previously reported in the literature as potential candidates for anodes in AMIBs \cite{zhu2018potential, liu2018few, wang2020ab}.
\begin{table}[hbt]
\small
  \caption{Adsorption properties of M atoms (M=Li, Na, K) in \ce{ABC-M1Si36Al18}. The table shows the adsorption energy $E_{ads}$ (in eV), the charge transfer $\rm{\Delta Q}$ (in $|e|$), and the equilibrium distances $d_{\rm M}$ (in {\AA}) between the adatom and the closest surface atom in the system.} 
  \label{table_energy}
  \begin{tabular*}{0.48\textwidth}{@{\extracolsep{\fill}}llccc}
    \hline
    &  &  &  &  \\ [-3mm] 
    Atom & Site & $E_{ads}$ & $\rm{\Delta Q}$ & $d_{\rm M}$ \\
    &  &  &  &  \\ [-3mm] 
    \hline
    &  &  &  &  \\ [-2mm] 
    Na & H & $-0.291$ & 0.811 & 1.905 \\
    &  &  &  &  \\ [-3mm] 
    Li & H & $-0.483$ & 0.859 & 1.177 \\
    &  &  &  &  \\ [-3mm] 
    Li & H$_2$ & $-0.470$ & 0.836 & 0.878 \\
    &  &  &  &  \\ [-3mm] 
    K & H & $-0.681$ & 0.803 & 2.423\\
    &  &  &  &  \\ [-3mm] 
    K & T$\rm _{Al}$ & $-0.481$ & 0.818 & 3.105\\
    &  &  &  &  \\ [-3mm] 
    K & T$\rm _{Si}$ & $-0.279$ & 0.826 & 3.072\\
     &  &  &  & \\ [-3mm] 
    \hline
  \end{tabular*}
\end{table}

Considering all adsorption sites investigated, the H is the most energetically favorable site to adsorb Li, Na, or K atoms with E$_{ads}$(K) ($-0.681$~eV) $<$ E$_{ads}$(Li) ($-0.483$~eV) $<$ E$_{ads}$(Na) ($-0.291$~eV). Similar behavior has been observed in pristine silicene monolayer, in which the H is the most favorable site for Li and Na adsorption \cite{tritsaris2013adsorption,mortazavi2016application}. The T sites are also suitable for adsorption of K atoms, as we found negative adsorption energies of $-0.481$~eV and $-0.279$~eV for T$_{\rm Al}$ and T$_{\rm Si}$, respectively. On the other hand, the adsorption energies for both Li and Na are positive at the T sites, indicating that they are not favorable for adsorption. Finally, the interstitial H$_{2}$ site is only energetically favorable for Li adsorption, with $E_{ads}=-0.470$~eV, while for Na and K ions we found positive values. The result regarding the H$_{2}$ site could be explained by considering the alkali metal atomic (ionic) radius and, hence, there is not enough space in this site to host Na or K ions. In particular, among all the \ce{ABC-M1Si36Al18} systems studied here, the ones related to K atoms seem to be the most promising to work in ion batteries, i.e., as an anode in KIBs, since the K ions can be adsorbed in several sites of the \ce{ABC-Si4Al2} trilayer host and, at the same time, they are the systems that present greater stability when compared to those that adsorb Li or Na, presenting the lowest values of $E_{ads}$.

In addition to the E$_{ads}$, table \ref{table_energy} lists the equilibrium distance ($d_M$) and the charge transfer ($\rm \Delta Q$). The $d_M$ is defined as the vertical distance (in the $z$-direction) between the adatom and the closest neighboring surface atom. We observed that $d_M$ increases with the increasing atomic (ionic) radius and mass of the adatoms, where $d_{\rm K}> d_{\rm Na}>d_{\rm Li}$ for the adsorption sites on the surface. The $d_M$ depends on the adsorption position, e.g., regarding T$_{\rm Al}$ and T$\rm _{Si}$ sites, the closest surface atoms are the Al and the Si atoms, respectively. Nevertheless, the closest surface atom is the Al for the H site, while it is the Si atom for the H$_2$ one inside the structure.
We also computed the $\rm \Delta Q$ from the ions toward the ABC-Si$_4$Al$_2$ host, using the Bader analysis \cite{henkelman2006fast, sanville2007improved, tang2009grid, yu2011accurate}. The values presented in table \ref{table_energy} are around +0.8 $|e|$, showing that the valence charge of the ions is almost fully transferred to the Si$_4$Al$_2$ host and, hence, indicating the ionization of the adatoms.

\subsection{Ion Diffusion Properties}

We then investigated the diffusion of the alkali metal atoms on ABC-Si$_4$Al$_2$ since the charge and discharge rates of the AMIBs are directly associated with the ion mobility \cite{shao2019recent}. Herein, we used the NEB method and a $3 \times 3 \times 1$ supercell,  in which two H adjacent sites were chosen as the initial and final configurations since it is the most energetically favorable site, i.e.,  it presents the largest adsorption energy value.

\begin{figure}[hbt]
\centering
\includegraphics[scale = 0.2, trim={0cm 0cm 0cm 0cm}, clip]{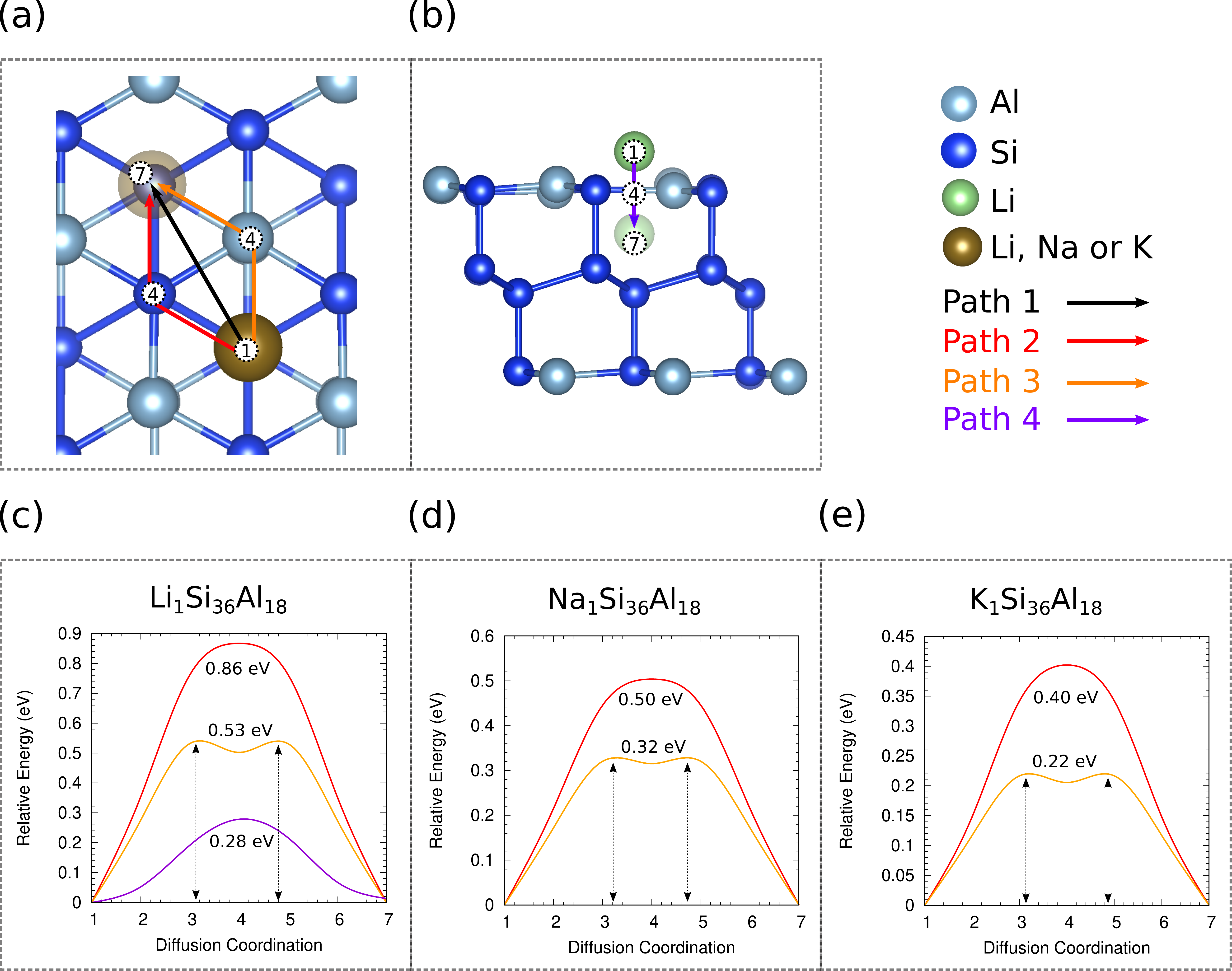} \quad
\caption{(a) Top and (b) side view of the ABC-Si$_4$Al$_2$ structure, indicating the binding sites (diffusion coordination 1, 4 and 7) and paths for Li/Na/K adatoms. The black, red, and orange arrows in (a), as well as the purple one in (b), represent the high-symmetry paths explored in this investigation (paths 1, 2, 3, and 4, respectively, as discussed in the text). The figure also shows the energy barriers associated with the diffusion of (c) Li; (d) Na; and (e) K; in ABC-Si$_4$Al$_2$. The  grey, blue, and brown spheres represent the Al, Si, and alkaline metal ions (Li, Na, K), respectively. Moreover, in (b) the green spheres represent the Li ion.}
\label{figure_3}
\end{figure}

Three different high-symmetry pathways on the \ce{ABC-Si4Al2} surface were considered, as illustrated in figure \ref{figure_3} (a). In Path 1, a single alkali atom moves from an H site to another H one passing through a B site. In Path 2, the atom passes through a T$_{\rm Si}$ site, while in Path 3, it passes above a T$_{\rm Al}$ one. Particularly, for the Li atom, we considered an additional path, moving it from the H to the H$_{2}$ site once the $E_{ads}$ is negative in the  H$_{2}$ site for the Li atom, which shows that it could be incorporated inside the structure. Figure \ref{figure_3} (b) illustrates this additional path (Path 4). At the end of NEB calculations, Path 1 converged toward Path 3, indicating that the ions diffuse to the top of the aluminum/silicon sites, and then to the other H site. An equivalent diffusion path was also observed in pristine silicene monolayer \cite{tritsaris2013adsorption}, which is consistent with what we previously observed regarding the alkali atoms moving from the B site to the T$_{\rm Al}$ one after accomplishing the structural optimization. 

The energy profiles along different paths are presented in figures \ref{figure_3} (c), (d), and (e) for Li, Na, and K ions, respectively. Comparing the same path, the calculated diffusion barriers ($E_b$) decrease as the atomic (ionic) radius and mass of the diffusing atoms increase: $E_b({\rm Li})>E_b({\rm Na})>E_b({\rm K})$. At this point, it is worth noting that the diffusion coordination 4 is the T$_{\rm Si}$ site on Path 2, the T$_{\rm Al}$ site on Path 3, and the H$_{1}$ site on Path 4. For Path 3, it can be observed in figures \ref{figure_3} (c), (d), and (e) that the peak is not at the diffusion coordination 4; instead, there is a  valley in that coordination that is due to the interaction between the alkali ions and the aluminum atom,  which was also observed in pristine silicene \cite{tritsaris2013adsorption, zhu2016silicene}.
The $E_b$ values regarding the ABC-Si$_4$Al$_2$ trilayer surface, Paths 2 and 3, are in the 0.53-0.86 eV, 0.32-0.50 eV, and 0.22-0.40 eV ranges for the systems with Li, Na, and K adatoms, respectively. 
For the Li atom, the lowest $E_b$ value of 0.28 eV is found in Path 4, suggesting that lithium atoms alloy with the ABC-Si$_4$Al$_2$ structure during the lithiation.

Table \ref{table_energy_barrier} presents the  Li, Na, and K  ions diffusion barriers values  in the ABC-Si$_4$Al$_2$ system, as compared with values reported in the literature for others candidate materials to serve as AMIBs anodes. The surface diffusion barriers $E_b$ corresponding to Path 3 for Li, Na, and K atoms are 0.53, 0.32, and 0.22 eV, respectively. These migration energy barriers values are close to the ones found in other silicon-based systems proposed as anode materials, such as AlSi$_3$ \cite{zhu2018potential}, Si$_3$C \cite{wang2020ab}, and Si$_2$BN \cite{shukla2017curious}, as shown in table \ref{table_energy_barrier}.

\begin{table}[hbt]
  \caption{\ Diffusion barrier energy ($E_b$), in eV, theoretical capacity ($C$), in mAh/g, and Open-Circuit Voltage (OCV), in V, for Li, Na, K ions in \ce{ABC-Si4Al2} trilayer system, as compared to results reported in the literature for other potential candidates to work as AMIB anodes. Herein, the $C$ and the OCV values are computed for the  Li$_{4}$Si$_4$Al$_2$, Na$_2$Si$_4$Al$_2$, and K$_2$Si$_4$Al$_2$ systems.}
  \label{table_energy_barrier}
  \scalebox{0.77}{\begin{tabular}{lllllllllllll}
    \hline
      &  &  &  &   &  &  &  &  &   &  &  &  \\[-2mm]
  Anode &  & \multicolumn{3}{c}{Diffusion barrier energy ($E_b$)} &  & \multicolumn{3}{c}{Theoretical Capacity  ($C$)} &  & \multicolumn{3}{c}{Open-Circuit Voltage (OCV)}  \\
    &  &  &  &   &  &  &  &  &   &  &  &  \\[-2.5mm]
Material &  & Li & Na & K  &  & Li & Na & K &   & Li & Na & K  \\
    \hline
    &  &  &  &   &  &  &  &  &   &  &  &  \\[-2mm]
 \ce{ABC-Si4Al2} & & 0.53-0.86 & 0.32-0.50 & 0.22-0.40 &   & 645 & 322  & 322 &   & 0.49 & 0.16  & 0.14 \\
  &  &  &  &   &  &  &  &  &   &  &  &  \\[-2mm]
Graphite & & 0.22-0.40 \cite{persson2010thermodynamic}&  &  &   & 372 \cite{schweidler2018volume} &   & 279 \cite{jian2015carbon} &   & 0.11 \cite{jing2013metallic} &   &   \\
 &  &  &  &   &  &  &  &  &   &  &  &  \\[-2mm]
 \ce{TiO2} &  & 0.35-0.65 \cite{lunell1997li, lindstrom1997li+, olson2006defect} & 0.52 \cite{lunell1997li}   &  &   & 200 \cite{lindstrom1997li+} &   & &   & 1.5-1.8 \cite{jing2013metallic} &   &   \\
  &  &  &  &   &  &  &  &  &   &  &  &  \\[-2mm]
 Silicene & & 0.23 \cite{tritsaris2013adsorption} & 0.16 \cite{zhu2016silicene} & 0.14 \cite{sahin2013adsorption}  &   & 954 \cite{tritsaris2013adsorption} & 954 \cite{zhu2016silicene}  & &   &  & 0.30-0.50 \cite{zhu2016silicene} &   \\
  &  &  &  &   &  &  &  &  &   &  &  &  \\[-2mm]
c-silicene &  &  &    &  &   & 721 \cite{liu2018few} &   & 180 \cite{sun2022reversible}  &   &  &  &  \\
 &  &  &  &   &  &  &  &  &   &  &  &  \\[-2mm]
 \ce{VS2} & & 0.22 \cite{jing2013metallic} & 0.62 \cite{putungan2016metallic} &  &   & 466 \cite{jing2013metallic} & 233 \cite{putungan2016metallic}  &   &   & 0.93 \cite{jing2013metallic} & 1.32 \cite{putungan2016metallic}  &    \\
  &  &  &  &   &  &  &  &  &   &  &  &  \\[-2mm]
\ce{AlSi3} \cite{zhu2018potential} & &  & 0.29   & 0.17  &   &  & 1928  & 964 &   &  &   &    \\
 &  &  &  &   &  &  &  &  &   &  &  &  \\[-2mm]
\ce{Si3C} \cite{wang2020ab} & & 0.47 & 0.34   & 0.18 &   & 1394 & 1115  & 836 &   & 0.58 & 0.50  & 0.71   \\
 &  &  &  &   &  &  &  &  &   &  &  &  \\[-2mm]
 \ce{Si2BN} \cite{shukla2017curious} & & 0.48 & 0.32   &  &   & 1158 & 933  & 836 &   & 0.46 & 0.29  &    \\
  &  &  &  &   &  &  &  &  &   &  &  &  \\[-2mm]
    \hline
    \end{tabular}}
   \end{table}

Our results indicate that the diffusion barrier of K ion on ABC-Si$_4$Al$_2$ is lower than those of Li and Na atoms, reinforcing which was previously observed that the K atom possesses more sites to be adsorbed, as well as presents the highest negative adsorption energies and equilibrium distances, when compared to the respective values of Li and Na atoms. Moreover, the diffusion barriers of the systems studied here are similar to those in anode materials based on graphite (0.22-0.40 eV) \cite{persson2010thermodynamic} and TiO$_2$ (0.35-0.65 eV) \cite{lunell1997li, lindstrom1997li+, olson2006defect}, implying that the ABC-Si$_4$Al$_2$ could be used as LIB, SIB, or KIB anodes.

To further compare the diffusion properties among those paths, we have obtained the diffusion constant (D) of an alkali metal atom, estimating it by the Arrhenius equation \cite{arrhenius1889dissociationswarme}:
\begin{equation}
    D \sim exp  \left(- \frac{E_b}{k_BT}\right),
    \label{eq_arrhenius}
\end{equation}
\noindent
where $E_b$ and $k_B$ are the diffusion barrier and Boltzmann’s constant, respectively, and $T$ is the environment temperature. Therefore, at room temperature, we estimate that the diffusion constant of Path 3, for a single Li, Na, and K, could be more than 10$^3$ times faster than Path 2. Moreover, regarding Path 3, the K mobility on ABC-Si$_4$Al$_2$ is more than 10$^5$ times faster than that of the Li ion and 49 times faster than that of Na one, being as fast as in commercial graphite anodes. Accordingly, we expect good high-rate performance for the ABC-Si$_4$Al$_2$ as the anode material, primarily in KIBs.

\subsection{Open-Circuit Voltage and Theoretical Capacity}

The open-circuit voltage (OCV) is the maximum voltage available for a battery, while the theoretical capacity ($C$) is the maximum amount of ions that can be stored in the host \cite{he2019density}. This section explores the adsorption energy ($E_{ads}$) as a function of the number of Li, Na, or K atoms adsorbed on \ce{ABC-Si4Al2}, in order to estimate the OCV and the $C$ by using equations \ref{eq_ocv} and \ref{eq_capacity}, respectively. 

\begin{figure}[b!]
\centering
\includegraphics[scale = 0.6, trim={0cm 0cm 0cm 0cm}, clip]{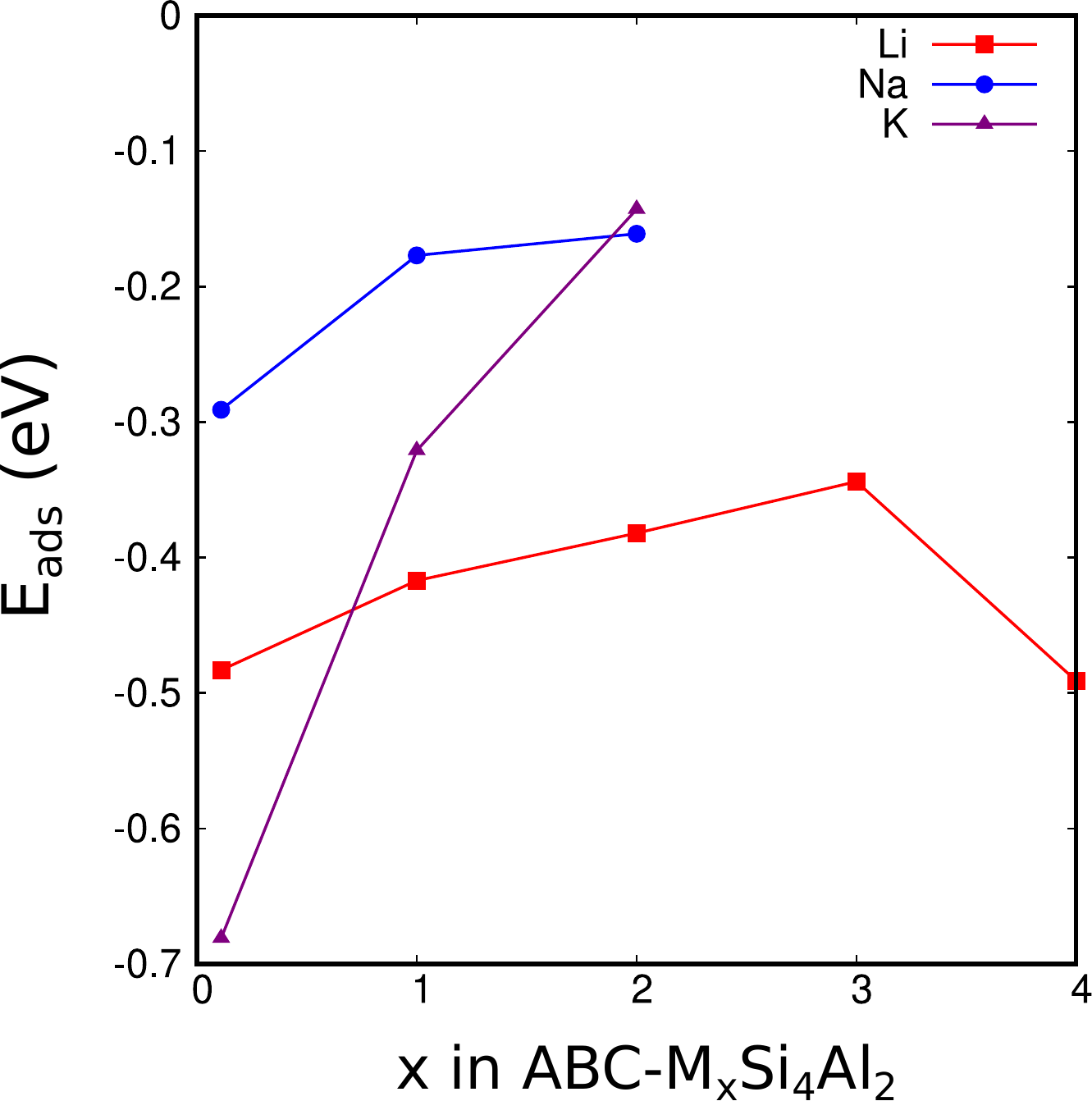} \quad
\caption{Adsorption energies ($E_{ads}$) as  function of M concentration (x) in \ce{ABC-M_{x}Si4Al2} (M = Li, Na, K).}
\label{figure_4}
\end{figure}

\begin{figure*}[htb]
\centering
\includegraphics[scale = 0.12, trim={0cm 0cm 0cm 0cm}, clip]{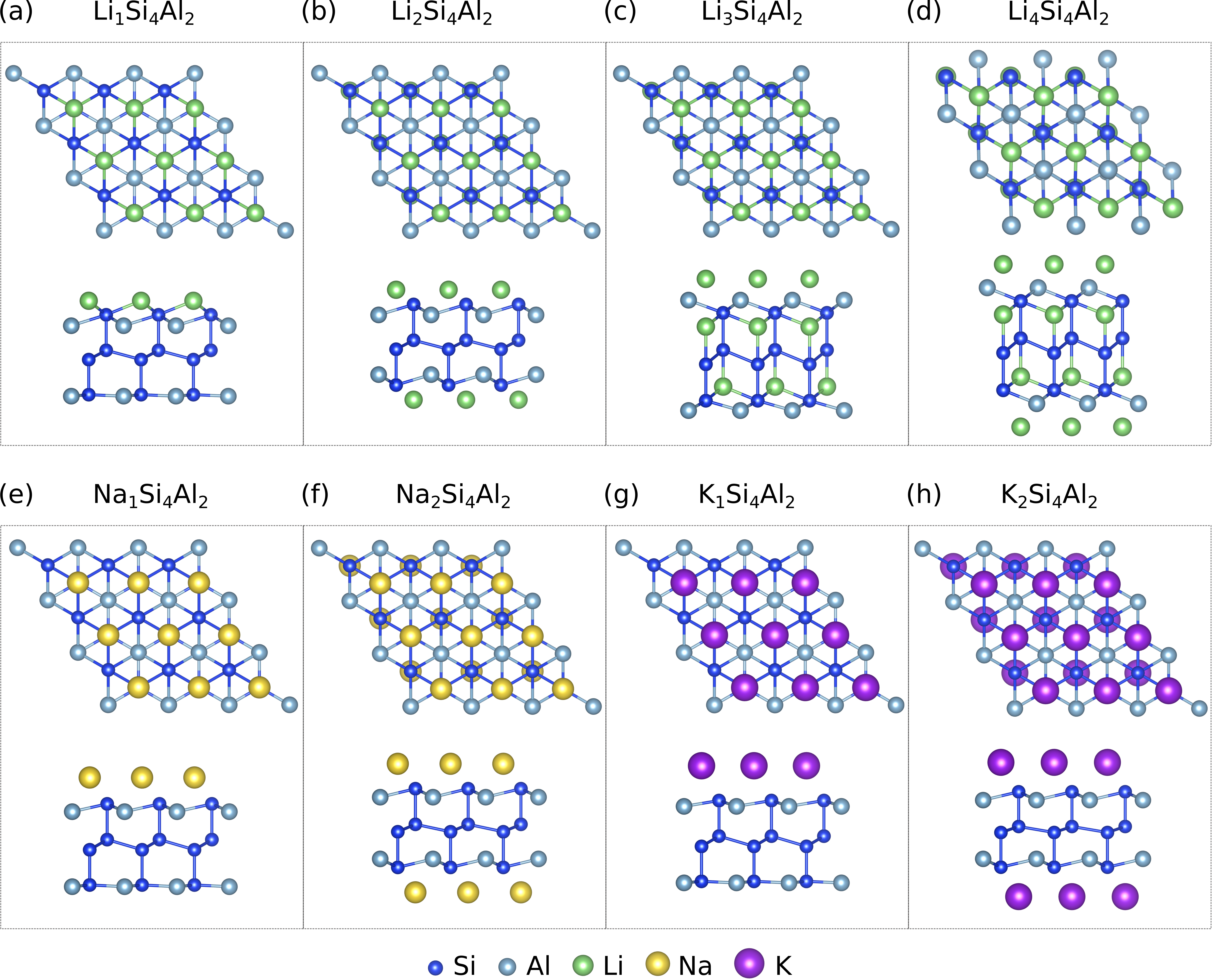} \quad
\caption{\ce{ABC-M_{x}Si4Al2} (M = Li, Na, K)  optimized structures for several alkaline metal concentration values (x).}
\label{figure_5}
\end{figure*}

Herein, we studied the \ce{ABC-Si4Al2} single- and double-side surface adsorptions, as well as the incorporation of Li atoms in the H$_2$ sites (figure \ref{figure_1}). Accordingly, we considered several configurations of \ce{ABC-Li_{x}Si4Al2}, \ce{ABC-Na_{y}Si4Al2}, and \ce{ABC-K_{y}Si4Al2} systems ($\rm{x = 0.11, 1, 2, 3, 4}$, and $\rm{y = 0.11, 1, 2}$) with the adatoms initially placed in the most favorable adsorption positions. Adsorption energies are presented in figure \ref{figure_4}, while the corresponding optimized structures are shown in figure \ref{figure_5} (a)-(h). It is worth mentioning that the first points in figure \ref{figure_4} represent the ${\rm x = 0.11}$ concentration and the $E_{ads}$ energy  of a single-side adsorption of M$_1$ alkali metal adatom in the H site of a 3$\times$3$\times$1 supercell (\ce{ABC-M1Si36Al18}), which was reported in the previous section.

Figure \ref{figure_4} shows that the adsorption energy continuously decreases as the M concentration increases, except for Li for $\rm{x}>3$. During ion adsorption, there are two important effects: the repulsion force between neighboring atoms and the structural changes. The repulsion force is mainly responsible for the adsorption energies of all ions decrease in the $0< {\rm x} < 3$ range, while the structural changes are dominant when ${\rm x}> 3$,  increasing the adsorption energy \cite{jiang2017boron}. Interestingly, the $E_{ads}$ for Li, Na, and K remained negative throughout the range explored here, indicating the \ce{ABC-Si4Al2} is a suitable anode material for Li-, Na-, or K-ion batteries. 

Figures \ref{figure_5} (a), (e), and (g) report the resulting configurations of single-side surface adsorption of M$_1$ adatom on the \ce{ABC-Si4Al2}, in which the H site was fully covered by Li, Na, or K atoms (\ce{ABC-M1Si4Al2}). Figures \ref{figure_5} (b), (f), and (h) show the double-side surface adsorption with the H site fully covered (\ce{ABC-M2Si4Al2}). For the Li-ion, we also have \ce{ABC-Li3Si4Al2} and \ce{ABC-Li4Si4Al2} systems that are shown in figures \ref{figure_5} (c) and (d), respectively. The most stable configuration of \ce{ABC-Li3Si4Al2} has all H$_2$ positions fully loaded and one surface side with all H sites fully covered by Li atoms.

Although it is possible to find several other phases at low concentrations by using different supercell sizes, we focused on the full loaded structure to estimate the OCV and the $C$. Accordingly, the OCV and $C$ were estimated for Li$_{4}$Si$_4$Al$_2$, Na$_2$Si$_4$Al$_2$, and K$_2$Si$_4$Al$_2$ systems, with the respective values presented in table \ref{table_energy_barrier}.  The OCV for Li-, Na-, and K-related systems are, respectively, 0.49 V, 0.16 V and  0.14 V, which are comparable with values in commercial anode materials, such as graphite (0.11 V) and \ce{TiO2} (1.5-1.8 V) \cite{jing2013metallic}, as well as other candidates for anodes, such as silicene (0.30-0.50 V) \cite{tritsaris2013adsorption, zhu2016silicene}, \ce{VS2} (0.93-1.32 V) \cite{jing2013metallic, putungan2016metallic}, \ce{Si3C} (0.50-0.71 V) \cite{wang2020ab}, and \ce{Si2BN} (0.29-0.46 V) \cite{shukla2017curious}. The low voltage values, which we found for \ce{ABC-Si4Al2}, are desirable characteristics for anode electrodes \cite{ma20212021}. 

The theoretical capacity of the \ce{ABC-Si4Al2} is 645 mAh/g for LIBs, and 322 mAh/g for both SIBs and KIBs, which are higher than the capacity in commercial anode materials, such as graphite (372 mAh/g for LIBs \cite{schweidler2018volume} and 279 mAh/g for KIBs \cite{jian2015carbon}), and \ce{TiO2} (200 mAh/g for LIBs \cite{lindstrom1997li+}). Moreover, the Li capacity on \ce{ABC-Si4Al2}  of 645 mAh/g is slightly smaller than an experimental value of 721 mAh/g found in pristine crystalline silicene as anode material for LIBs \cite{liu2018few}. Remarkably, the K capacity on \ce{ABC-Si4Al2} of 322 mAh/g is higher than 180 mAh/g when pristine crystalline silicene was experimentally investigated as anode material for KIBs \cite{sun2022reversible}. The crystalline silicene presents an experimental thickness of $\approx$ 6.5 {\AA} and lattice parameter of $\approx$ 3.4 {\AA} \cite{liu2018few, sun2022reversible}. This system appears to be structurally similar to the few-layer pristine silicene nanosheets made of silicene trilayer vertically combined. Indeed, we recently investigated two stacking configurations (AA$'$A and ABC) of pristine silicene trilayers bonded covalently, and the values obtained for the thickness and the lattice parameter were $\approx$ 7.0 {\AA} and 3.858 {\AA}, respectively \cite{ipaves2022functionalized}. Due to the aluminum doping, the \ce{ABC-Si4Al2} trilayer exhibits a smaller thickness of 6.0 {\AA} and a larger lattice parameter of 4.143 {\AA} \cite{ipaves2022functionalized}.  This lattice expansion could be the reason for the increase of the K adsorption, i.e., the higher capacity of K on \ce{ABC-Si4Al2}, as compared to crystalline silicene. Therefore, our results indicate that ABC-Si$_4$Al$_2$ has a superior theoretical capacity for alkali metal ion storage, primarily for KIBs.

\section{Concluding Remarks}

In summary, the \ce{ABC-Si4Al2} was systematically investigated as a potential anode material for Li, Na, and K ion batteries. Our previous investigation has shown that this structure is dynamically and mechanically stable, and herein the simulations have shown that \ce{ABC-Si4Al2} is also thermodynamically stable up to 600 K. We explored the adsorption and diffusion properties of alkali metal ions (Li, Na, K) on \ce{ABC-Si4Al2}, finding large adsorption energies and low diffusion barriers values. Moreover, the \ce{ABC-Si4Al2} provides good theoretical capacity and low open-circuit voltage. The lowest OCV value was obtained for the K adatom and the highest theoretical capacity was found for Li one. Our estimated surface diffusion barriers, OCV, and specific capacity are comparable to the respective values in commercial anode materials, such as graphite and \ce{TiO2}, and other candidate materials, such as pristine silicene. We also found a slightly smaller capacity for LIBs and a higher one for KIBs as compared with the experimental values of pristine crystalline silicene. Those results suggest that doped few-layer silicene is a potential material for anode batteries, with a good high-rate performance and theoretical capacity for alkali metal storage (Li, Na, K), primarily for SIBs and KIBs. Furthermore, the functionalization of few-layer silicon-based systems could be a route to developing novel electrodes for AMIBs.








\bibliography{achemso-demo}

\end{document}